\documentclass[useAMS,usenatbib]{mn2e}
\usepackage{graphicx}
\newcommand{\pq}{\ensuremath{P_Q}}
\newcommand{\pu}{\ensuremath{P_U}}

\title[A successful search for Barbarians in the Watsonia family]{A successful
search for hidden Barbarians in the Watsonia asteroid family
\thanks{Based on observations made with ESO Telescopes at the La Silla-Paranal Observatory
under programme ID 091.C-0027}}
\author[A. Cellino et al.]{A. Cellino$^{1}$\thanks{E-mail:
cellino@oato.inaf.it (AC); sba@arm.ac.uk (SB); ptanga@oca.eu
(PT); bojan@matf.bg.ac.rs (BN); mdelbo@oca.eu (MDB)},S.
Bagnulo$^{2}$, P. Tanga$^{3}$, B. Novakovi\' c$^{4}$ and M.
Delb\`o$^{3}$\\
$^{1}$INAF - Osservatorio Astrofisico di Torino, Pino Torinese 10025, Italy\\
$^{2}$Armagh Observatory, College Hill, Armagh BT61 9DG, UK\\
$^{3}$University of Nice Sophia Antipolis, CNRS, Observatoire de la
C\^ote d'Azur, Laboratoire Cassiop\'ee, France\\
$^{4}$Department of Astronomy, Faculty of Mathematics, University of
Belgrade, 11000 Belgrade, Serbia}

\begin{document}

\date{Accepted 2013 December 20. Received 2013 December 20; in original form 2013 October 15}

\pagerange{\pageref{firstpage}--\pageref{lastpage}} \pubyear{2013}

\maketitle

\label{firstpage}

\begin{abstract}
Barbarians, so named after the prototype of this class (234)
Barbara, are a rare class of asteroids exhibiting anomalous
polarimetric properties. Their very distinctive feature is that they
show negative polarization at relatively large phase-angles, where
all ``normal'' asteroids show positive polarization. The origin of
the Barbarian phenomenon is unclear, but it seems to be correlated
with the presence of anomalous abundances of spinel, a mineral
usually associated with the so-called Calcium Aluminum-rich
inclusions (CAIs) on meteorites. Since CAIs are samples of the
oldest solid matter identified in our solar system, Barbarians are
very interesting targets for investigations. Inspired by the fact
that some of the few known Barbarians are members of, or very close
to the dynamical family of Watsonia, we have checked whether this
family is a major repository of Barbarians, in order to obtain some
hints about their possible collisional origin. We have measured the
linear polarization of a sample of nine asteroids which are members
of the Watsonia family within the phase-angle range
$17\degr-21\degr$. We found that seven of them exhibit the peculiar
Barbarian polarization signature, and we conclude that the Watsonia
family is a repository of Barbarian asteroids. The new Barbarians
identified in our analysis will be important to confirm the possible
link between the Barbarian phenomenon and the presence of spinel on
the surface.
\end{abstract}


\section{Introduction}
\label{Intro} The degree of linear polarization of sunlight
scattered by an asteroid toward an observer depends on the
phase-angle, namely the angle between the asteroid - Sun and the
asteroid - observer directions. The phase-polarization curves of all
atmosphereless bodies of the solar system exhibit qualitatively
similar trends, but their detailed features vary according to the
specific properties (including primarily the geometric albedo) of
individual surfaces. In the phase-angle range $\sim 0\degr \div
20\degr$, asteroids exhibit the so-called branch of \textit{negative
polarization}, in which, in contrast to what is expected from simple
single Rayleigh-scattering or Fresnel-reflection mechanisms, the
plane of linear polarization turns out to be parallel to the plane
of scattering (the plane including the Sun, the target and the
observer). The plane of linear polarization becomes perpendicular to
the scattering plane, a situation commonly described as
\textit{positive polarization}, at phase angle values larger than
the so-called \textit{inversion angle}, which is generally around
$20\degr$.

A few years ago, \citet{Celetal06} discovered a class of asteroids
exhibiting peculiar phase-polarization curves, characterized by a
very unusual extent of the negative polarization branch, with an
inversion angle around $30\degr$, much larger than the values
commonly displayed by most objects. Since the prototype of this
class is the asteroid (234) Barbara, these objects have been since
then commonly known as \textit{Barbarians}. Only half a dozen
Barbarians are known today: (234) Barbara, (172) Baucis, (236)
Honoria, (387) Aquitania, (679) Pax, and (980) Anacostia
\citep{Celetal06,Giletal08,MasCel09}.

The polarimetric properties of the Barbarians are fairly unexpected.
The observed large negative polarization branch is not predicted by
theoretical models of light scattering, but in fairly special
situations, including surfaces composed of very regularly-shaped
particles (spheres, crystals) or surfaces having considerable
microscopic optical inhomogeneity \citep{Shkuratov94}. Although
Barbarians are certainly uncommon, they do exist, and the
interpretation of their polarization features may lead to important
advances in our understanding of both light-scattering phenomena,
and of the origin and evolution of these objects. Potential
explanations range from peculiar surface composition and/or texture,
to the possible presence of anomalous properties at macroscopic
scales due the presence of large concavities associated with big
impact craters \citep{Celetal06}. For instance, (234) Barbara has a
very long rotation period, which might be the effect of a big
collision. \citet{Deletal09} suggested that (234) Barbara could have
a surface characterised by large-scale craters. This is confirmed by
an analysis of still unpublished occultation data by one of us (PT).

In terms of taxonomy based on spectro-photometric data, all known
Barbarians are classified as members of a few unusual classes,
including $L$, $Ld$, and (in only one case) $K$. (234) Barbara
itself is an $Ld$ asteroid \citep[here we use the taxonomic
classification of][]{BusBin02}.  However, there are $L$-class
asteroids which are ``normal" objects not exhibiting the Barbarian
properties. This fact seems to rule out a direct relationship
between taxonomic class (based on the reflectance spectrum) and
polarimetric properties. On the other hand, $L$, $Ld$ and $K$
classes are located, in a Principal Component Analysis plane, along
adjacent locations, which although non-overlapping, seem to
represent some kind of continuous spectral alteration surrounding
the most common $S$ class complex. The fact that the six known
Barbarians identified so far belong all to one of these three
classes suggests that surface composition could be responsible for
their polarimetric properties. Even more important, two $L$-class
Barbarians, (387) Aquitania and (980) Anacostia, exhibit very
similar reflectance spectra, both sharing the rare property of
displaying the spectral signature of the spinel mineral
\citep*{Buretal92}. Actually, it was exactly the fact that (980)
Anacostia was found to be a Barbarian that led \citet{MasCel09} to
observe polarimetrically (387) Aquitania, and to discover that also
this object shares the same polarimetric behaviour.

Spinel ([Fe,Mg]Al$_{2}$O$_4$) is a mineral characterized by
indistinct cleavage and conchoidal, or uneven fracture properties.
In terms of optical properties, the MgAl$_{2}$O$_4$ form of spinel
has a fairly high refractive index ($n=1.72$), which becomes even
higher in the spinel variety having a high iron content (hercynite)
\citep[$\sim 1.8$, i.e., much above the values characterizing the
most common silicates present on asteroid
surfaces,][]{Sunshine2008}. Spinel is an important component of
Calcium Aluminum-rich inclusions (CAI) found in all kinds of
chondritic meteorites. CAIs are refractory compounds which are
thought to be among the first minerals to have condensed in the
proto-solar nebula. They are the oldest samples of solid matter
known in our solar system, and they are used to establish the epoch
of its formation \citep{Sunshine2008}. In terms of spectroscopic
properties, spinel is characterized by the absence (or extreme
weakness) of absorption bands around 1\,$\mu$m, and by the presence
of a strong absorption band at 2\,$\mu$m. \citet{Sunshine2008}
concluded that, to model the available near-IR spectra of
spinel-rich asteroids, it is necessary to assume abundances of the
order of up to 30\,\% of CAI material on the surface. This extreme
abundance, which causes a high refractive index, might also be
responsible for the anomalous polarization properties. Such high CAI
abundances have never been found in meteorite on Earth (so far, the
richest CAI abundance, found on CV3 meteorites, is about 10\,\%).
Therefore, \citet{Sunshine2008} conclude that spinel-rich asteroids
``might be more ancient than any known sample in our meteorite
collection, making them prime candidates for sample return"
missions.

Many interesting questions are certainly open. Which processes are
involved in the onset of physical mechanisms which produce the
Barbarian polarimetric behaviour? Are Barbarians really among the
oldest objects accreted in our solar system? If so, why are they the
only one class of primitive objects being characterized by an
anomalous polarimetric behaviour?  Why are they so rare? Are they
unusually weak against collisions and fragmentation? Do
space-weathering phenomena affect their polarimetric properties? Are
the taxonomic classifications of some $L$, $Ld$, $K$ objects
possibly wrong (in such a way that all Barbarians might be member of
a unique class and not spread among three of them)? It is clear that
an important pre-requisite to improve our understanding of these
objects, and to draw from them some possible inferences about the
origin, composition and subsequent evolution of the planetesimals
orbiting the Sun at the epoch of planetary formation, is to find new
members of the Barbarian class, but where to look for them?

An aid to a Barbarian search comes from the fact, recently confirmed
by \citet*{Novetal11}, that the spinel-bearing Barbarian (980)
Anacostia belongs to a family of high-inclination asteroids. This
family is named Watsonia from its lowest-numbered member (729)
Watsonia. (980) Anacostia belongs to a small grouping which is
included in the Watsonia family and merges with it at larger values
of mutual distances between the members. In other words, (980)
Anacostia belongs to a distinct sub-group of the Watsonia family,
but the possible independence of this sub-group from the rest of the
family is highly uncertain and questionable \citep{Novetal11}.
Another known Barbarian, (387) Aquitania, though not being a member
of the Watsonia family, is also located in close vicinity in the
space of proper orbital elements (see Section\,\ref{Results}).
Finally, a member of the Watsonia family, asteroid (599) Luisa, is
not a known Barbarian (no published polarimetric measurements are
available for it), but it is known to be one of the few spinel-rich
asteroids identified so far (thus sharing some common properties
with both Anacostia and Aquitania).

One should also note that the distribution of albedos of the members
of the Watsonia family observed at thermal IR wavelengths by the
WISE satellite \citep{Masieroetal11} is strongly peaked around
values between $0.10$ and $0.15$ with only a few likely interlopers,
suggesting that the family is not a statistical fluke\footnote{We
examined the WISE albedo distribution of the Watsonia family by
using the $MP^3C$ web facility available at URL
http://mp3c.oca.eu/}. A possible common collisional origin of these
asteroids opens new perspectives for the search of new Barbarians
and for the interpretation of their properties. Asteroid families
are the outcomes of fragmentation of single parent bodies disrupted
by catastrophic collisions. Therefore, if (980) Anacostia, and
possibly also (387) Aquitania, were issued from the disruption of a
common parent body exhibiting the rare properties which produce the
Barbarian polarization phenomenon, it would be natural to expect
that also the other, still not observed members of the Watsonia
family should be found to be Barbarians. Moreover, among the
Watsonia family members at least one, (599) Luisa, is a known
spinel-rich asteroid, like Anacostia and Aquitania. Finally, WISE
albedo values in the same range of those of Watsonia family members
have also been derived for most Barbarian asteroids known so far,
namely 234, 172, 236, 679 and 980.

\section{New observations}
\begin{table*}
\caption{\label{Tab_Observations} Polarimetry of nine asteroids of
the Watsonia family in the special Bessell $R$ band. $\pq$ and $\pu$
are the reduced Stokes parameters measured in a reference system
such that $\pq$ is the flux perpendicular to the plane
Sun-Object-Earth (the scattering plane) minus the flux parallel to
that plane, divided by the sum of the two fluxes. It is therefore
identical to the parameter often indicated as $Pr$ in asteroid
polarimetry studies. The last column lists, when available, the
albedo value according to WISE thermal IR measurements}
\begin{center}
\begin{tabular}{ccrrcr@{\,$\pm$\,}lr@{\,$\pm$\,}lc}
\hline \hline Date                             & Time (UT) &
\multicolumn{1}{c}{Exp}          & \multicolumn{1}{c}{Object} &
Phase angle                      & \multicolumn{2}{c}{\pq} &
\multicolumn{2}{c}{\pu}          & WISE \\
\multicolumn{1}{c}{(yyyy mm dd)} & \multicolumn{1}{c}{(hh:mm)} &
(sec)                            & \multicolumn{1}{c}{} &
\multicolumn{1}{c}{(DEG)}        & \multicolumn{2}{c}{(\%)} &
\multicolumn{2}{c}{(\%)}         & albedo \\
\hline
            &       &      &          &       & \multicolumn{4}{c}{}     \\
 2013 07 05 & 23:41 &  480 & 5492     & 18.79 & $-$1.14 & 0.10 &    0.05 & 0.10 & $0.14\pm0.02$\\
 2013 07 29 & 01:44 &  960 &          & 18.31 & $-$1.01 & 0.09 &    0.00 & 0.09 \\[2mm]
 2013 06 03 & 09:40 & 2000 & 42365    & 23.30 & $-$0.83 & 0.15 & $-$0.06 & 0.15 & $0.17\pm0.02$\\
 2013 08 03 & 09:08 &  960 &          & 18.55 & $-$1.73 & 0.12 & $-$0.06 & 0.12 \\[2mm]

 2013 07 12 & 23:50 &  960 & 56233    & 17.83 & $-$1.07 & 0.16 & $-$0.09 & 0.16 & $0.19\pm0.03$\\
 2013 08 05 & 01:03 & 2200 &          & 19.31 & $-$1.09 & 0.12 & $-$0.07 & 0.12 \\[2mm]

 2013 07 30 & 00:38 & 2000 & 106059   & 18.30 & $-$1.06 & 0.30 & $-$0.15 & 0.31 \\
 2013 08 04 & 01:09 & 4000 &          & 18.82 & $-$0.94 & 0.14 &    0.00 & 0.14 \\
 2013 08 28 & 01:24 & 4800 &          & 19.64 & $-$0.84 & 0.20 & $-$0.09 & 0.20 \\[2mm]

 2013 07 06 & 01:33 & 1400 & 106061   & 20.21 & $-$0.94 & 0.11 &    0.06 & 0.11 \\
 2013 08 09 & 02:46 & 4000 &          & 23.97 & $-$0.57 & 0.12 & $-$0.05 & 0.12 \\[2mm]

 2013 07 06 & 02:05 &  960 & 144854   & 21.30 & $-$0.81 & 0.12 &    0.13 & 0.12 \\
 2013 08 05 & 02:06 & 4000 &          & 24.19 & $-$0.15 & 0.12 & $-$0.13 & 0.13 \\[2mm]

 2013 08 13 & 06:28 & 4000 & 236408   & 18.31 & $-$0.97 & 0.15 &    0.22 & 0.15 & $0.14\pm0.04$\\[2mm]

 2013 07 07 & 02:40 & 3440 & 247356   & 19.97 &    0.10 & 0.15 & $-$0.33 & 0.15 \\[2mm]

 2013 04 17 & 08:58 & 4800 & 320971   & 23.78 &    0.11 & 0.31 & $-$0.06 & 0.31 \\
 2013 06 03 & 07:08 & 3440 &          & 20.53 & $-$0.03 & 0.22 & $-$0.06 & 0.21 \\[2mm]
\hline
\end{tabular}
\end{center}
\end{table*}
\begin{figure}
 \includegraphics[width=75mm]{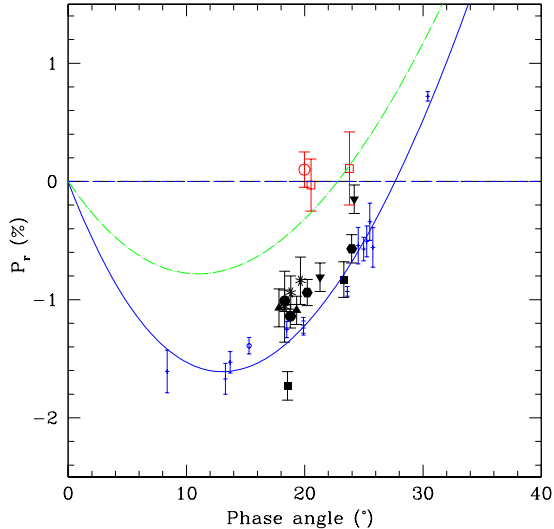}
 \caption{Phase - polarization data (in $R$ light) for the targets of the present
 investigation, compared with the polarization curves (in $V$ light) of the
 (234) Barbara, and of (12) Victoria, a large $L$-class asteroid which does not
 exhibit the Barbarian
 behaviour. Black symbols: the seven targets exhibiting the
 Barbarian polarimetric behaviour; red symbols: our two targets that
 display a "normal" polarimetric behaviour; small blue symbols and blue curve:
 available data for (234) Barbara \citep{MasCel09}, and the corresponding best-fit
 curve using the linear-exponential relation
 $ P_r = A [e^{\alpha/B}-1] + C \cdot \alpha$, where $\alpha$ is the phase angle in
 degrees; dashed, green curve: the best-fit curve for the $L$-class asteroid (12) Victoria
 (for the sake of clearness, individual observations of this asteroid are not shown).}
 \label{Figure1}
\end{figure}
Our target list includes nine objects that are members of the
Watsonia family (not limited to the Anacostia sub-group) according
to \citet{Novetal11}. These objects are listed in
Table~\ref{Tab_Observations}. We note that a more recent family
search whose results are publicly available at the AstDys web
site\footnote{http://hamilton.dm.unipi.it/astdys/index.php?pc=5}
also identified a Watsonia family, but using a more conservative
value of the critical level of mutual distances between family
members to be adopted to define the family. As a consequence, the
AstDys version of the Watsonia family has a smaller membership list,
and does not include our target objects 42365, 144854, 247356 and
320971.

We observed our targets using the VLT FORS2 instrument
\citep{Appetal98} in imaging polarimetric mode, and obtained $17$
broadband linear polarization measurements in the $R$ special filter
from April to September 2013. The choice of an $R$ filter (instead
of the standard $V$ filter traditionally used in many asteroid
polarimetric observations) was dictated by the need of improving the
S/N ratio. Polarimetric measurements were performed with the
retarder wave-plate at all positions between 0\degr\ and 157.5\degr,
at 22.5\degr\ steps. For each observation, the exposure time
cumulated over all exposures varied from 8\,min (for 5492) to 1\,h
and 20\,min (for 320971).  Data were then treated as explained in
\citet{Bagetal11}, and our measurements are reported adopting as a
reference direction the perpendicular to the great circle passing
through the object and the Sun. This way, $\pq$ represents the flux
perpendicular to the plane Sun-Object-Earth (the scattering plane)
minus the flux parallel to that plane, divided by the sum of these
fluxes. It is therefore identical to the parameter indicated as
$P_r$ in many asteroid polarimetric studies. In these conditions,
for symmetry reasons, $\pu$ values are expected to be zero, and
inspection of their values confirms that this is the case for our
observations.

At small phase angles ($\ll 20\degr$), all asteroids exhibit
negative polarization, whereas at phase angles $\ga 20\degr$, nearly
all asteroids exhibit positive polarization. By contrast, at phase
angle $\simeq 20\degr$, Barbarians still exhibit a relatively strong
negative polarization ($\sim -1$\,\%).
Therefore, to identify Barbarians, we decided to measure the
polarization at phase angles $\ge 17\degr$, and establish whether
the measured polarization plane is found to be parallel or
perpendicular to the scattering plane.

\section{Results}
\label{Results} The results of our polarimetric measurements are
given in Table~\ref{Tab_Observations} and shown in
Figure~\ref{Figure1}, in which a comparison is also made with the
phase-polarization curve of (234) Barbara and (12) Victoria, a big,
non-Barbarian $L$-class asteroid. In the Figure each target is
represented by a different symbol, and some targets have more than
one measurement. In the observed phase-angle range, seven of our
targets show a polarization value $\sim -1$\,\%, consistent with the
value exhibited by (234) Barbara. In fact, (234) Barbara exhibits
marginally higher polarization values (in absolute value) than our
targets, possibly due to the different filter in which the
observations were performed.

The striking result of our observing campaign is that seven out of
nine asteroids of our target list are Barbarians.  The two
exceptions are 247356 and 320971. However, we know \emph{a priori}
that all family member lists are expected to include some fraction
of random interlopers \citep{Miglio95}, then we conclude that both
non-Barbarian objects in our target list may be family interlopers.
For instance, asteroid 320971 is listed as a Watsonian member by
\citet{Novetal11}, but it is not included among the Watsonia members
identified in the new AstDys list of asteroid families. On the other
hand, three other targets not included in the AstDys member list,
but present in the \citet{Novetal11} member list, are found to be
Barbarians. We conclude therefore that, apart from some details
concerning family membership, the Watsonia family is an important
repository of Barbarian asteroids. This is the immediate and most
important result of our investigation. We also note that the number
of Watsonia Barbarians identified in our observing campaign is
larger than the whole sample of Barbarians previously known. We also
note that, except in case of 236408 and 247356 (only a single
measurement each) the resulting  $P_U$ values are always well
consistent with zero.

\section{Discussion and future work}
Our results confirm once again that asteroid polarimetry can provide
an important contribution to asteroid taxonomy, as noted in the past
by several authors \citep[see, for instance,][]{Penttilaetal05}.

Several problems, however, are now open, and deserve further
theoretical and observational efforts.

The first issue to be addressed is the relationship between the
Barbarian polarimetric features and the unusual amount of spinel
measured via spectroscopy in some of the known Barbarians. In other
words, the new Barbarians that we have found in our investigation
must be spectroscopically observed in the visible and near-IR
wavelengths in order to check whether they exhibit the spinel
features. We do believe that this will be the case, since we already
know three objects that are both spinel-rich and Barbarians (234
Barbara, 387 Aquitania and 980 Anacostia), ad we know also that the
Watsonia family includes at least one other spinel-rich asteroid
(599 Luisa) \citep{Sunshine2008}, but we clearly need confirmation
from new observations.

Similarly, we need to search for the Barbarian polarimetric feature
in other spinel-rich asteroids. In particular, we are interested not
only in 599 Luisa, but also in the Henan family, which is known to
include at least three spinel-rich asteroids \citep{Sunshine2008}.

Next, we need to understand why the Barbarian and the anomalous
spinel abundance phenomena are so rare. Apparently, only a handful of
Barbarian parent bodies existed, and only their disruption into many
fragments made it possible to identify today larger numbers of these
objects. In principle, one might wonder whether the unusual properties
displayed by these objects might also be in some (still obscure) way a
consequence of the collisional events themselves.

Another open problem is the origin of (387) Aquitania. This asteroid
is not included in any family list, including both the most recent
classification available at the AstDys site, and the \citet{Novetal11}
classification of high-$I$ families.  Yet, this object is quite close,
in terms of orbital elements, to (980) Anacostia and the rest of the
Watsonia family. The values of proper elements for (387) Aquitania and
(980) Anacostia (available in the AstDys database) are
$a = 2.73916$, $e = 0.23025$, $\sin I = 0.28245$, for (387)
Aquitania, and $a = 2.74102$, $e = 0.13972$, $\sin I = 0.29805$ for
(980) Anacostia, where $a$ is the proper semi-major axis, $e$ the
proper eccentricity, and $I$ is the proper inclination. The only one
relevant difference, which prevents any family search to include
them in an acceptable group, is the proper eccentricity.
One should conclude that there is no relation between (387)
Aquitania and (980) Anacostia, and that the similarity of their
orbital semi-major axes and inclinations are just a coincidence.
Moreover, both asteroids display stable orbits in terms of
characteristic Lyapunov exponents, and any possible evolution due to
non-gravitational forces, (e.g.\ the Yarkovsky effect) seem also
very unlikely, because both asteroids are fairly large, with sizes
of the order of $80 \div 100$ km according to thermal radiometry
data. Finally, the collision which could have produced a family
including two fragments of this size had to be extremely energetic,
and should have produced a huge family, of which there is no
evidence today.

On the other hand, Aquitania and Anacostia exhibit too many relevant
similarities in their physical properties,
which suggests that they may come from a unique parent body
fragmented by a collision. A very tentative scenario is that, in a
very early epoch, a violent collision destroyed a big parent body
which was the progenitor of the Barbarians we see today in this
zone. Most of the first-generation fragments produced by the event
are now gone due to gravitational perturbations and Yarkovsky-driven
drift in semi-major axis. The current Watsonia family could be the
product of a more recent disruption of one of the original fragments
of the first event. This region could still include several
first-generation fragments, preferentially the biggest ones which,
due to their size, experienced no or extremely weak Yarkovsky
evolution, like (387) Aquitania and (980) Anacostia. Some
gravitational perturbations could have also played some role. If one
looks at the $a$ - $e$ or $a$ - $\sin I$ plots making use of the
facility available at the AstDys
site\footnote{http://hamilton.dm.unipi.it/astdys2/Plot/}, it is easy
to see that the Watsonia family is crossed, in semi-major axis, by a
couple of thin resonances, one of which is located close to the
value of $a$ of both (387) Aquitania and (980) Anacostia. So, we
cannot rule out entirely the idea that some limited evolution in
eccentricity of these objects could be due to these perturbation
effects and the two objects might have been originally closer in the
proper element space. We stress again that this kind of scenario is
only eminently speculative and deserves further analysis before it
can be accepted as a plausible one.

We point out that a so large relative fraction of observed members
of the Watsonia family turning out to be Barbarians has a further
implications to be taken into account. If the parent body was
differentiated, or had some gradient of composition as a function of
depth, with the presence of CAI material limited to some thin shell,
then the spinel signature would be present only in a minority of the
family members.
This is in disagreement with what we observe, indicating that the
anomalous composition of the parent body was not limited to some
thin region. In conclusion, we are not looking simply at some kind
of surface process (like alteration by space weathering or impacts),
but at the properties displayed by asteroids originated from
different depths inside a single parent body which was anomalously
spinel-rich in most of its volume (this can be even more true if the
Watsonia family is a secondary event derived by a first-generation
destruction of a larger body.)

The issues of Section~\ref{Intro} are still open and deserve careful
attention.  The immediate result of our investigation is that we
have now a much larger sample of Barbarians than before at disposal
for future physical investigations (including photometric,
spectroscopic, and polarimetric campaigns.)  At the same time, our
results definitively confirm the family of Watsonia as a sizeable
group of Barbarians having a common collisional origin, and open new
perspectives for the development of models in order to try and sort
out in a coherent scenario several pieces of observational evidence
which appear now to form a complicated puzzle.

\section*{Acknowledgments}
Part of this work was supported by the COST Action MP1104
\emph{Polarization as a tool to study the Solar System and beyond}.
In particular, funding for Short Term Scientific Mission COST-
STSM-MP1104-12750 of AC at the Armagh Observatory is kindly
acknowledged. The work of B.N. has been supported by the Ministry of
Education and Science of Serbia, under the Project 176011. A careful
review by K. Muinonen helped to improve the paper.

\label{lastpage}

\end{document}